\documentclass[rnote]{aa}

\usepackage{graphicx}
\usepackage{txfonts}
\usepackage{natbib}
\usepackage{longtable}
\usepackage{lscape}
\usepackage{color}

\begin{document}
 
\title{Allsky NICER and NICEST extinction maps based on the 2MASS near-infrared survey
\thanks{The maps are available in electronic form at the CDS via anonymous ftp to 
cdsarc.u-strasbg.fr (130.79.128.5) or via http://cdsweb.u-strasbg.fr/cgi-bin/qcat?J/A+A/.
The maps can also be found at http://www.interstellarmedium.org/Extinction.}
}

\titlerunning{Allsky NICER and NICEST extinction maps}

\author{M.     Juvela\inst{1}
\and
        J.     Montillaud\inst{2},
}

\institute{
Department of Physics, P.O.Box 64, FI-00014, University of Helsinki,
Finland, {\em mika.juvela@helsinki.fi}
\and
Institut UTINAM, CNRS UMR 6213, OSU THETA, Universit\'e de Franche-Comt\'e, 
41 bis avenue de l'Observatoire, 25000 Besan\c{c}on, France
}

\authorrunning{M. Juvela et al.}

\date{Received September 15, 1996; accepted March 16, 1997}

\abstract { 
Extinction remains one of the most reliable methods of measuring column 
density of nearby Galactic interstellar clouds. The current and ongoing
near-infrared surveys enable the mapping of extinction over large sky areas.
} 
{
We produce allsky extinction maps using the 2MASS near-infrared survey.
}
{
We use the NICER and NICEST methods to convert the near-infrared colour
excesses to extinction estimates. The results are presented in Healpix
format at the resolutions of 3.0, 4.5, and 12.0 arcmin. 
}
{
The main results of this study are the calculated J-band extinction maps. The comparison
with earlier large-scale extinction mappings shows good correspondence but also
demonstrates the presence of resolution-dependent bias. A large fraction of the bias can
be corrected by using the NICEST method. 
}
{
For individual regions, best extinction estimates are obtained by careful
analysis of the local stellar population and the use of the highest
resolution afforded by the stellar density. However, the uniform allsky maps
should still be useful for many global studies and as the first step into the
investigation of individual clouds.
}


\maketitle

\section{Introduction}

Extinction measurements are one of the main methods by which the column
density of interstellar clouds can be measured. With the development of
near-infrared (NIR) observations, it has become possible to extend the
studies into dense clouds. The usual range accessed by NIR observations is
1--20 magnitudes of visual extinction, $A_{\rm V}$. However, with dedicated
deep observations, even higher values can be measured. This also makes NIR studies
relevant for studies of star formation from the formation of molecular
clouds to the structure of individual dense cores.

In the optical regime, extinction has been mapped using the star-counting method
\citep[see, e.g.][]{Dobashi2005_DSS}. The same technique can be extended to NIR
wavelengths, but it is more common to use the colour excesses of individual stars
\citep[e.g.][and references below]{Cambresy1997_DENIS}. This is based on two
observations. First, the dispersion of the intrinsic colours of the stars is relatively
small at NIR wavelengths \citep[e.g.][]{Lombardi2001_NICER, Cambresy2002_2MASS,
Davenport2014}. Secondly, the variations of the NIR extinction curve are also small
\citep[e.g.][]{Cardelli1989, Wang2014}. Thus, the comparison of the observed NIR colours
and the expected intrinsic colour gives a good estimate of the extinction. To bring the
uncertainty down to a few tenths of visual magnitude, the data need to average over ten
or more stars. Furthermore, to take into account the variations of stellar colours over
the sky, the intrinsic colour is estimated using a nearby extinction-free area. Although
each star gives an $A_{\rm V}$ estimate for a very narrow line of sight (LOS), the final
resolution is determined by the stellar density. With the 2MASS survey
\citep{Skrutskie2006}, a resolution of $\sim$1 arcmin can be reached but this degrades
rapidly towards high Galactic latitudes.

The extinction of several large areas has already been mapped using the 2MASS survey.
These include the Polaris Flare \citep{Cambresy2001}, the Pipe nebula
\citep{Lombardi2006_Pipe}, Ophiuchus and Lupus \citep{Lombardi2008_Oph}, Taurus
\citep{Padoan_2002_Taurus, Lombardi2010_Taurus}, Orion \citep{Lombardi2011_Orion,
Lombardi2014}, and Corona Australis \citep{Alves2014_CrA}. Most studies have used the
optimised multifrequency methods the near-infrared color excess revised (NICER)
\citep{Lombardi2001_NICER} or NICEST \citep{Lombardi2009_NICEST} to combine the
information of the $J-H$ and $H-K$ colours. Depending on the region and its local
density of background stars, the spatial resolution of these maps varies from one arcmin
to a few arcmin. In \citet{Cambresy2002_2MASS}, instead of using a fixed spatial
resolution, the resolution was varied based on the local stellar density, thus keeping
the signal-to-noise ratio (S/N) approximately constant.

The first all-sky maps of NIR extinction were presented by
\citet{Dobashi2011_2MASS}, including an extensive catalogue of dark
clouds detected in extinction. \citet{Dobashi2013_Besancon} included a
further modification where the Besan\c{c}on model of the Galactic
stellar populations was used to correct estimates of stellar intrinsic
colours. These papers included extinction maps based on the J-H and
H-K colours separately, but no all-sky NICER and NICEST maps have been
published so far. 

In this paper we describe the calculation of all-sky NIR extinction maps with the NICER
and NICEST methods. The structure of the paper is the following. In
Sect.~\ref{sect:methods} we describe the methods, including the calculation of the
intrinsic stellar colours. In Sect. \ref{sect:data} we describe the pre-filtering of
2MASS catalogue data. The results and comparison with some previously presented
extinction maps are given in Sect.~\ref{sect:results} before a short discussion of the
findings in Sect.~\ref{sect:discussion}. The final conclusions are listed in 
Sect.~\ref{sect:conclusions}.

\section{Methods}  \label{sect:methods}

We use two methods, NICER \citep{Lombardi2001_NICER} and NICEST
\citep{Lombardi2009_NICEST}, to calculate J-band extinctions based on 2MASS
data. Both methods are based on the comparison between the observed NIR
colours and the reference colours that would correspond to zero extinction.
We use the reference colours defined in Sect.\ref{sect:reference}.  Any
additional extinction of intervening dust reddens the observed colours, in
our case the J-H and H-K colours. In the (J-H, H-K) plane the change is
determined by the amount of extinction and the shape of the extinction
curve. Because of the dispersion in intrinsic stellar colours, both methods
require the averaging of the extinction estimates of many individual stars.

NICER technique seeks to combine two (or more) colour measurements in an optimal way.
It is a generalisation of the earlier Near-Infrared Colour Excess (NICE) method that
deals with single NIR colours \citep{Lada1994_NICE}. In NICER the reference colours are
described by their average values, $<J-H>$ and $<H-K>$, and their covariance matrix. The
dispersion of the values should be dominated by the scatter of the intrinsic colours,
not by the photometric errors of individual stars. The method first calculates
extinction estimates for each star. It solves the least squares problem taking into
account both the dispersion of the reference colours and the uncertainty of the colour
measurements of the individual stars. In practice, the
distribution of intrinsic colours must be assumed to be well characterised using just a covariance
matrix and, furthermore, that the same extinction law is applicable to all stars. Once
the estimates of individual stars are known, the final extinction values are obtained as
spatial averages calculated with a Gaussian smoothing kernel. Because of the scatter of
the intrinsic colours, there should be at least $\sim$10 stars within the area covered
by the kernel. In addition to weighted averaging, the NICER method includes the
so-called sigma-clipping procedure: individual stars are not included in the average if
their extinction estimates deviate more than $k_{\sigma} \sigma_{A}$ from the average.
Here $k_{\sigma}$ is a sigma-clipping threshold and $\sigma_{A}$ the standard deviation
of the extinction estimates of individual stars. The procedure is repeated until no more
stars get eliminated. This is useful, for example, to reduce the contamination by
foreground stars. For details of the implementation, see \citet{Lombardi2001_NICER}.

In the presence of column density variations, the estimates of NICE and
NICER are biased. If the extinction varies within the smoothing kernel, more
stars are seen through the lower extinction parts and, therefore, the
average extinction is underestimated. The problem can be significant at high
column densities, if column density variations are large. This applies
equally in the case of steep large-scale gradients and in the case of
complex structures not resolved by the chosen smoothing kernel.
The NICEST method adds a correction for the bias \citep{Lombardi2009_NICEST}. This
involves adjusting the weights of individual stars so that stars with higher extinction
estimates have larger influence. Furthermore, a correction that depends on the slope of
the cumulative star counts and on the uncertainties in the extinction of individual
stars is applied to the final extinction estimate. This ensures that the result remains
unbiased even when the averages are calculated using only a few stars. However, the bias
correction is more efficient if averages are calculated over more stars. NICEST cannot
solve the problem of such column density peaks where no background stars are observed.

In our implementations of NICER and NICEST we follow closely the descriptions
given in \citet{Lombardi2001_NICER} and \citet{Lombardi2009_NICEST}. In
practice, calculations are performed using Healpix pixelisation, smoothing
kernels with FWHM equal to 3.0, 4.5, and 12.0 arcmin, and the reference
colours described in Sect.~\ref{sect:reference}-\ref{sect:br}. NICEST also
requires an estimate of the slope $\alpha$ of the stellar density vs.
magnitude, $\rho \sim 10^{\alpha \times m}$. We use a single value of 0.31
(H-band) \citep[see also][]{Lombardi2009_NICEST, Dobashi2013_Besancon}. When
calculated directly from the 2MASS catalogue, $\alpha$ shows a latitude
dependence, rising from $\sim$0.1 at Galactic poles up to $\sim$0.34 in the
plane (median values over constant latitudes). However, the value
$\alpha$=0.31 is valid to within $\pm$0.05 units for all latitudes up to
$|b|\sim 30\degr$. This covers most regions where the NICEST method could be
useful. We assume the parameterised extinction curve of \citet{Cardelli1989}
($R_{\rm V}$=3.1) with $A_{\rm J}/E(J-H)$=2.77 and $A_{\rm J}/E(H-K)$=4.44.
These values are not sensitive to assumptions of the selective extinction
$R_{\rm V}$. For $R_{\rm V}$=3.1 our A$_{\rm J}$ values can be converted to
$A_{\rm V}$ by multiplying them with 3.55.

In addition to the normal sigma-clipping procedure, we will calculate a
few variations of the NICER maps, using stars selected based on their
extinctions relative to other stars within the smoothing kernel. Maps Q1-Q4
correspond to different quartiles. Thus Q1 is calculated with 25\% of stars
with the smallest $A_{J}$ values and Q4 with 25\% of the largest $A_{J}$
values. Finally, Q12 and Q34 employ stars below and above the median value,
respectively.

Large differences are not expected between the new NICER maps and those presented in
\citet{Dobashi2011_2MASS} and \citet{Dobashi2013_Besancon}. Those earlier maps were also
based on the J-H and H-K colours of 2MASS stars, the $A_{\rm V}$ values being calculated
with the NICE method \citep{Lada1994_NICE}. However, with the NICER method the
reliability of the extinction estimates is improved by the optimal combination of the
J-H and H-K colours, the expected gain being a factor of two in the variance of the
extinction estimates \citep{Lombardi2009_NICEST}. All stars detected in at least two
bands are used. Compared to NICE and NICER estimates, NICEST maps should differ because
of their smaller bias (and somewhat larger noise). This may be important in dense clouds
because, due to the large size of the smoothing kernel, averages are calculated over a
wide range of column densities. Without bias correction, the extinction values can be
underestimated by tens of per cent or even more \citep{Lombardi2009_NICEST,
Juvela2014a}.

\section{Input data}  \label{sect:data}

\subsection{2MASS stars}

The input data consists of the 2003 data release of the 2MASS point source catalogue
(PSC) \citep{Skrutskie2006}. We started by rearranging the data into files, one for each
of the pixels of the Healpix NSIDE=16 pixelisation \citep[see][]{Gorski2005}. This
results in 3072 files where each file includes all the point sources falling within the
corresponding healpix pixel (some 3.7 degree in size) or outside the pixel but closer
than $\sim$20.0$\arcmin$ of its boundaries. This allows each input file to be used
independently to calculate an extinction map for an area corresponding to one healpix
NSIDE=16 pixel. When spatial smoothing is done with a Gaussian beams with
FWHM=12$\arcmin$ and kernel is placed at the boundary of a NSIDE=16 pixel, the data
still cover an area with more than 99.995\% of the convolution kernel.

In the file conversion, we carry over the coordinates, magnitude, magnitude
errors, and the flags $ph\_qual$, $rd\_flag$, and $mp\_flg$. We drop all
sources with $mp\_flg$ set (sources associated with known solar system
objects) and further require $rd\_flag$ equal to 1, 2, or 3 and $ph\_qual$
equal to $A$, $B$, or $C$ (best quality detections with reliable photometry).
However, we do not drop sources where only one band, either J or K, fails the
above selection. This was done under the assumption that the other colour
(J-H or H-K) still carries useful information and the influence of the more
uncertain magnitude measurement is weighted down by ensuring it has a very
large error estimates. We also reject extended sources with either of the
flags $ext\_key$ or $gal\_contam$ set. After these criteria we are left with
a total of 409 million sources, some 87\% of the full PSC catalogue.

\subsection{The reference colours} \label{sect:reference}

In addition to the apparent magnitudes of stars behind a column of dust,
extinction estimation requires information of the average intrinsic stellar
colours and their covariances. For individual clouds, these are often
calculated using the stars in a nearby low-extinction region. 
This corresponds to an idealised scenario where a foreground cloud is seen against
a constant background. In this case, even if the chosen reference region is not
completely free of extinction, the extinction relative to the reference region would be
correctly traced. When the reference region contains significant amounts of extinction,
the estimated intrinsic colours become redder and also our estimate of the dispersion of
the intrinsic colours is likely to increase. Nevertheless, if the background is
constant, a nearby reference region provides a good approximation of the average colour
of the stars as their radiation reaches the foreground cloud. In a general case the
concept of a ``background star'' is not well defined and depends on what part of the
medium along the LOS is being studied.

\begin{figure}
\includegraphics[width=8.8cm]{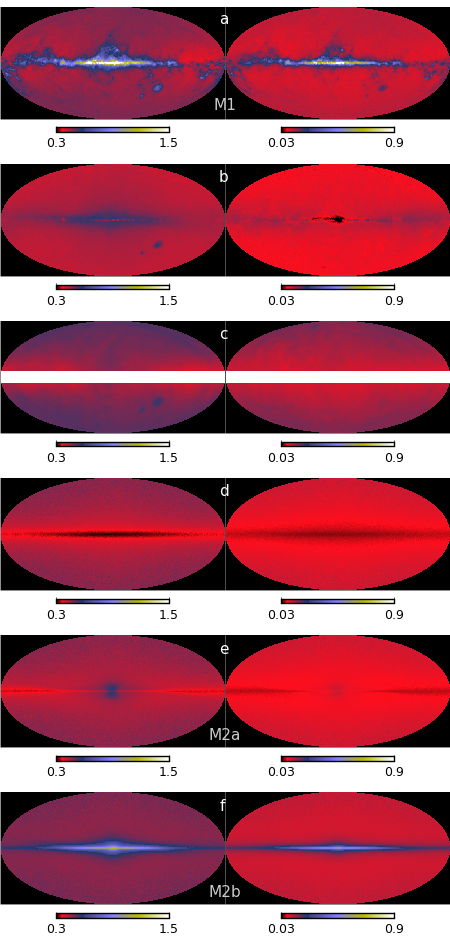} 
\caption{
Six versions of all-sky maps of the average J-H (left frames) and H-K (right frames)
colours. The first three are based on observed 2MASS sources (frames a-c) and the last
three (frames d-f) on stars simulated with the Besan\c{c}on model. See text for details.
The labels M1, M2a, and M2b refer to extinction maps where the corresponding data
are used as reference colours.
}
\label{fig:plot_colours}
\end{figure}

For an all-sky map the procedure is even less straightforward. We need
reference colours that are a compromise between the actual intrinsic stellar
colours and the apparent stellar colours for the radiation reaching nearby
clouds. The difference is relevant mostly at low Galactic latitudes. The
reference colours should also be continuous over the whole sky to avoid
discontinuities in the extinction maps. We have examined colours extracted
directly from the 2MASS data and colours obtained from stars simulated with
the Besan\c{c}on model. Figure~\ref{fig:plot_colours} shows six versions of the
maps of J-H and H-K colours.

In the first maps (frame a), the values are calculated as direct averages of
all stars falling inside a given NSIDE=256 Healpix pixel (pixel size $\sim
13.7\arcmin$). These values are affected by whatever foreground extinction
the stars have and, especially in the plane, the colours are strongly
reddened. These are thus not a particularly good reference for extinction
calculations, with the possible exception of compact clumps whose extinction
is estimated relative to the local extended background. Nevertheless, we
will use the average values at high-latitudes $|b|>60\degr$ as the most
basic reference to be used in extinction calculations.

The second maps (frame b) is calculated also on NSIDE=256 resolution, in each pixel
dropping 30\% of the stars located in the highest extinction areas. For this procedure,
the extinction values were read from a NICER extinction map produced with the
high-latitude reference colours (see above); only relative extinctions at scales much
below one degree are needed. 
The 30\% cut-off is arbitrary but is used to mask regions that are strongly
affected by extinction (mainly by distinct nearby clouds). The results 
depend only weakly on the exact value of this threshold.
The colours of the remaining stars are further corrected using the $E{\rm (B-V)}$ map of
\citet{Schlegel1998}, assuming an extinction curve with $R_{\rm V}=3.1$. In the
correction all extinction is assumed to reside between the stars and us. This results in
over-correction, especially near the plane. The plots of the resulting J-H and H-K
colours are included in Fig.~\ref{fig:plot_colours}b only for reference and they are not
used in actual extinction calculations.

The third maps are calculated at NSIDE=64 resolution, dropping all stars with
extinction exceeding 1\,mag  in Fig.~\ref{fig:plot_colours}a.
This means that for many pixels near the
Galactic plane, no estimates can be derived. For the remaining stars below
1\,mag, we apply now a small correction that is again derived from
\citet{Schlegel1998} $E{\rm (B-V)}$ map, this time assuming that half of the
extinction resides between the stars and us. The J-H and H-K maps are shown
in Fig.~\ref{fig:plot_colours}c, where we have masked all pixels
$|b|<8\degr$. As in the frame $b$, the correction has removed most of the
column density related structures at higher latitudes. Again, these maps are
included only for illustration and are not used any further.

The last three pairs of maps in Figure~\ref{fig:plot_colours} are based on
stars simulated using the Besan\c{c}on model \citep{Robin2014}. In frames d 
and e we have used only stars within a
distance of either 2.0\,kpc or 8.0\,kpc, respectively. At high latitudes,
these are identical and close to the values derived from 2MASS stars. On the
other hand, both J-H and H-K now decrease towards the centre of the Galactic
plane. When the sample is extended to 8\,kpc distances, stars in the central
bulge increase J-H and H-K towards the Galactic centre. In these plots stars
are not affected by any extinction. Because the extinction maps will be used
mostly to examine nearby structures, it would seem necessary to assume that
very distant stars are affected by at least some diffuse extinction. Thus,
the light from the stars behind a nearby cloud should already be reddened
when it reaches our cloud. The selected procedure is described in
Sect.~\ref{sect:br}

\subsection{Selected reference colours} \label{sect:br}

Besan\c{c}on model provides description of the intrinsic colours of stars as a function
of Galactic location. When we examine a given region on the sky, it is likely to contain
structures at different distances. Conversely, if we are examining a nearby cloud, the
``background stars'' can be at very different distances and they may be affected by
varying amounts of extinction not related to our cloud. If we define the colour of
background stars based on the radiation reaching a nearby cloud, extinction further
along the LOS changes the reference colours and their dispersion. This
affects the error estimates of extinction and also directly the results of the NICEST
method.

The problem cannot be fully addressed without information about the distribution of
interstellar medium along the LOS. If we relied directly on the intrinsic colours of
Besan\c{c}on stars, in the plane the values would be affected by distant stars that in
reality are never observed.  We attempt to make a crude correction using the following
procedure. In the mid-plane we assume a diffuse extinction of $A_{\rm
V}$=1.0\,mag\,kpc$^{-1}$. This is lower than the canonical value of
$\sim$1.8\,mag\,kpc$^{-1}$ which includes the contribution of dense clouds. To estimate
the dependence on Galactic height, we first examined the all-sky $E(B-V)$ map of
\citep{Planck2014_XI}, which traces the total amount of material along the LOS.
We calculated the average profile of this total extinction as a function of Galactic
height, using median values calculated over all Galactic longitudes. Assuming that the
profile is mostly determined by relatively nearby regions, we modelled the Galactic
plane as a slab with single profile $f(z)$ as a function of Galactic height $z$.
When $f$ is assumed to be an exponential function, we obtain a scale height of $\sim
210$\,pc. Thus, this 3D density distribution reproduces the latitude
dependence of LOS extinction as seen in the Planck $E(B-V)$ map. The scale height is
higher than the typical values quoted for the ISM. This is possibly an effect of nearby
cloud complexes (Taurus-Auriga, Orion, etc.) that should not be counted in our model of
diffuse extinction. Therefore, we finally adopted a steeper exponential function with a
scale height 150\,pc. As mentioned above, the absolute scaling is determined by using at
$z=0$\,pc a value of 1.0\,mag\,kpc$^{-1}$.

In Fig.~\ref{fig:plot_colours}, the last frame $f$ shows the colours that are calculated
using Besan\c{c}on model, including the diffuse extinction as defined above. Based on
its distance and Galactic latitude, the NIR magnitudes of each star are increased with
the corresponding amount, using $R_{\rm V}=3.1$ extinction curve to calculate the NIR
extinction. Stars falling beyond detection limits of 15.8, 15.1, and 14.3\,mag in J, H,
and K bands, respectively, are dropped. In the plane, the J-H and H-K colours are now
higher because of the diffuse extinction. This should be a better reference when
estimating the extinction of nearby clouds. However, we again emphasise that without a
full 3D image of the interstellar medium, it is not possible to calculate the
``correct'' reference colours that would result in correct extinction estimates in
absolute terms. The final extinction maps should not be used, for example, to estimate
the distribution of interstellar medium as a function of latitude. In the plane the zero
point of the extinction scale is essentially arbitrary.

In the following, we calculate extinction maps using three versions of reference
colours. The first extinction maps, called M1, correspond to the average statistics of
2MASS stars at Galactic latitudes $|b|>60\degr$. The reference colours of the M1 maps
thus represent the mean properties over the high-latitude sky as seen in
Fig.~\ref{fig:plot_colours}a. This serves as a useful point of reference when we test
the sensitivity of the extinction values on the assumptions concerning the reference
colours.
The extinction maps M2a and M2b employ the reference colours shown in
Figs.~\ref{fig:plot_colours}e-f. Map M2a uses the average colours of all simulated stars
up to a distance of 8\,kpc. For version M2b these colours are modified by diffuse
extinction that is calculated from the model described above, using the distance
distribution of simulated stars. 
Thus, M2a has higher and M2b lower absolute values of extinction and the difference should
be most noticeable at low latitudes.
To calculate the reference colours, the Besan\c{c}on model was used to create stellar
catalogues with over 879 million stars over the full sky. The average colours and their
covariances are calculated for each NSIDE=256 pixel (size $\sim$13.7$\arcmin$), finally
smoothing the parameter maps with a Gaussian beam with FWHM equal to one degree. 

\section{Results}  \label{sect:results}

\begin{figure}
\includegraphics[width=8.8cm]{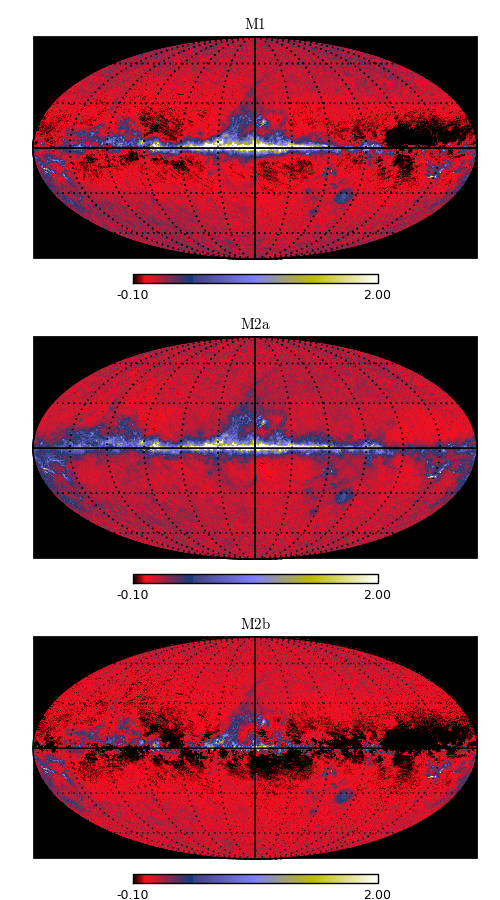}
\caption{
M1, M2a, and M2b versions of NICER extinction maps. The colour scale is
linear with respect to $A_{\rm J}$.
}
\label{fig:allsky_M14ab}
\end{figure}

Figure~\ref{fig:allsky_M14ab} shows the extinction maps calculated with the
NICER method and using the M1, M2a, and M2b versions of reference colours.

The main difference is in the latitude dependence of the large scale
distribution. The M1 and M2a versions are relatively similar although M2a
values increase more consistently towards the Galactic plane. This is also
the expected behaviour, especially if the M2a reference colours
underestimate the effective reddening of the ``background stars''. On the
other hand, M2b map has values closer to zero but large areas at
mid-latitudes have negative values. Again, this is consistent with the M2b
reference colours being too red because of the model we used for the
extended extinction.

\begin{figure}
\includegraphics[width=8.8cm]{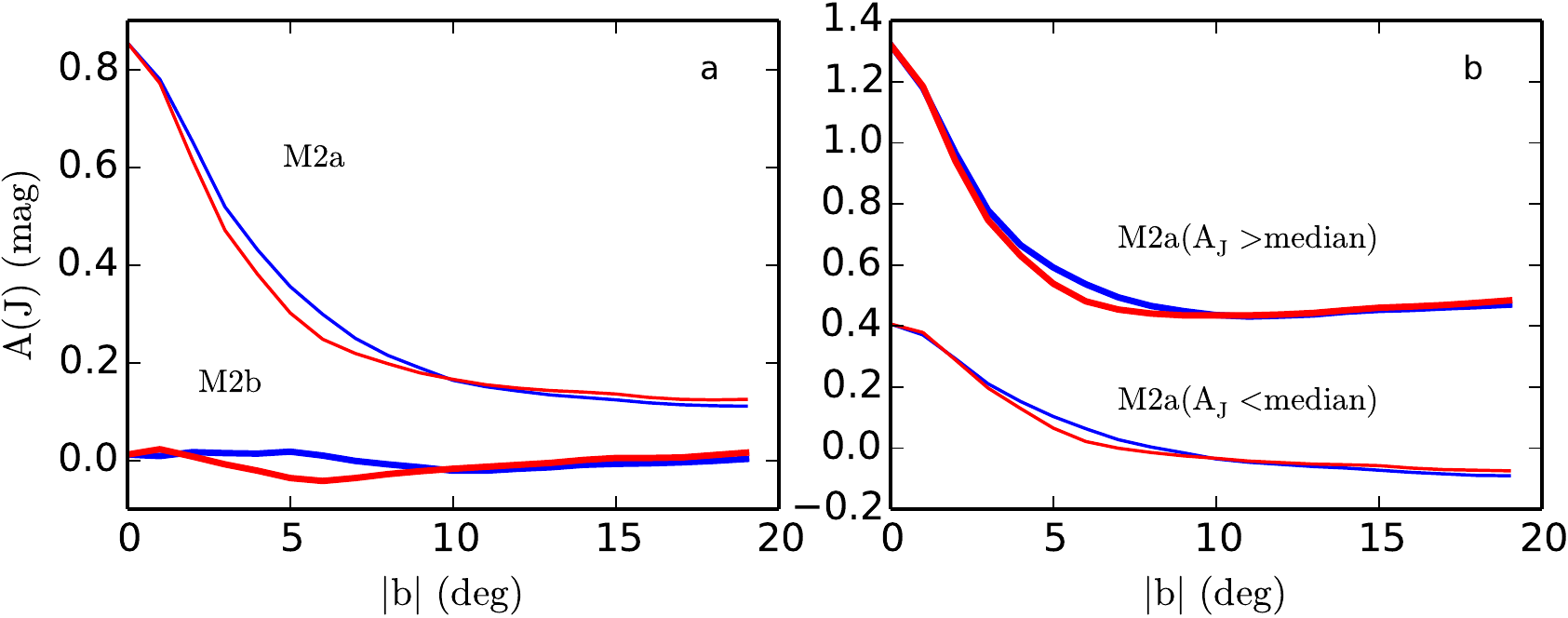} 
\caption{
Comparison of median NICER $A_{\rm J}$ profiles as function of Galactic
latitude. The left frame compares M2a (thin lines) and M2b maps (thick
lines) for the profiles at positive (blue lines) and negative (red lines)
Galactic latitudes.
Frame $b$ shows comparison of M2a maps constructed using those stars that
have values below (thin lines) or above (thick lines) the median $A_{\rm J}$
of all stars within the smoothing kernel. 
}
\label{fig:M2a_M2b}
\end{figure}

The latitude dependence of $A_{\rm J}$ is better visible in
Fig.~\ref{fig:M2a_M2b}, which plots the median extinction against the
latitude. The left frame shows how the extinction in the M2a maps rises
towards the Galactic mid-plane, the profile being almost identical on the
two hemispheres. On the other hand, the latitude profiles of M2b version are
flat with values close to zero. In other words, in this case the reference
colours are very close to the stellar colours reddened by the average
extinction at each latitude. In principle, M2b represents the correct
reference colours for a very nearby object whose extinction is seen above
the average Galactic background. Because of the strong spatial variations of
column densities, this is almost never precisely the case.

Figure~\ref{fig:M2a_M2b}b shows the also variations of the M2a maps where
map is constructed using only the stars that are either below or above the
median value of $A_{\rm J}$ of all stars within the smoothing kernel. The
maps are at $3\arcmin$ resolution. At high latitudes, the values are mostly
noise and the selection of either low extinction or high extinction stars
results in a significant bias, $\Delta A_{\rm J}\sim 0.5$\,mag. Near the
plane, the steeper rise of the $A_{\rm J}>median$ map should also have
physical reasons related to unresolved high column density structures.
Furthermore, when stars and extincting structures are mixed along LOS,
$A_{\rm J}<median$ favours nearby clouds, resulting in larger scale height.

\begin{figure}
\includegraphics[width=8.8cm]{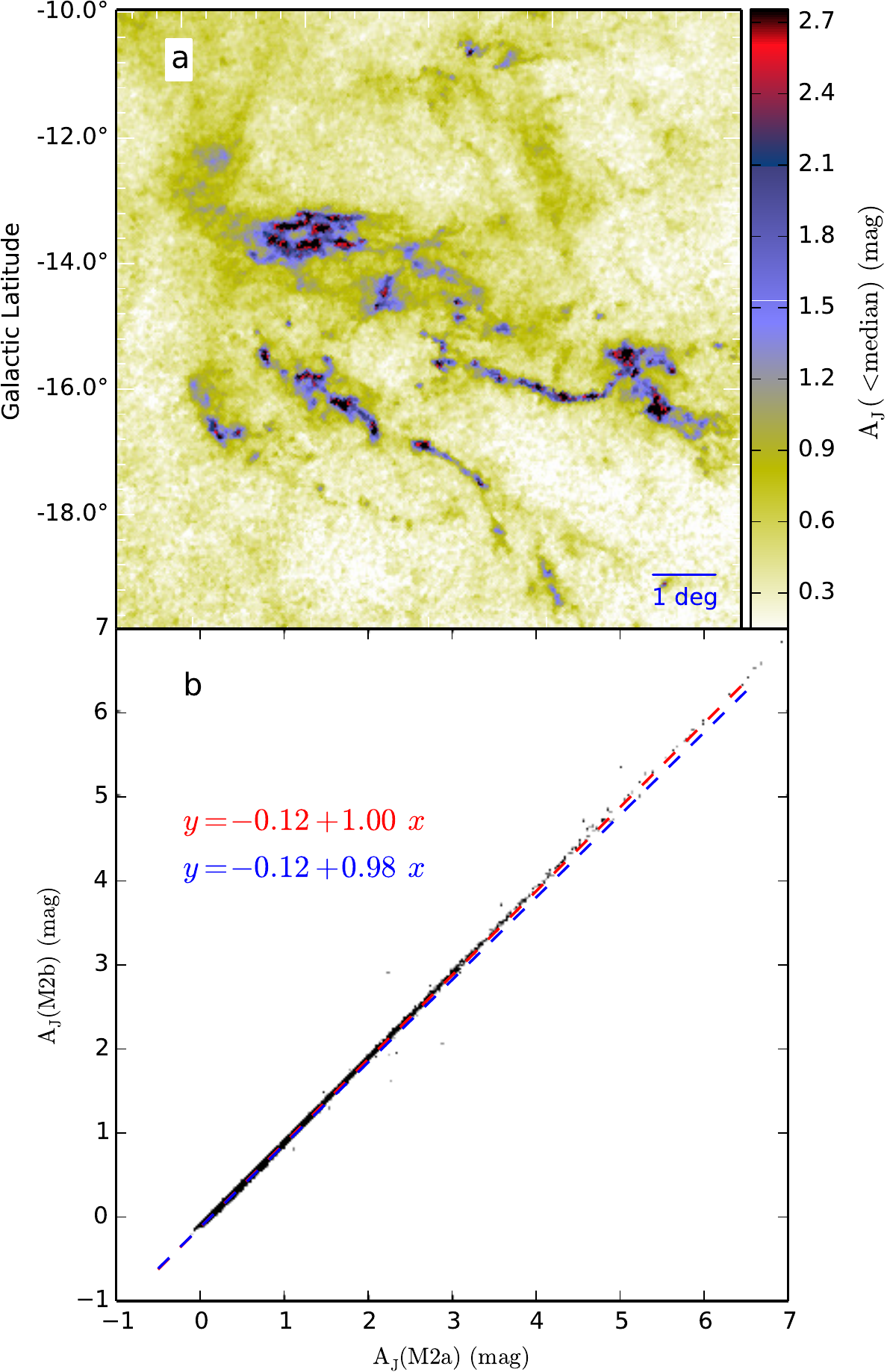} 
\caption{
The extinction map of Taurus and the correlation between the M2a and M2b
estimates within the field. The scatter plot includes all pixels of the
image (black dots). The red dashed line is the least squares fit and the
blue line is a with slope set equal to one.
}
\label{fig:M2ab_Taurus}
\end{figure}

All extinction map versions (M1, M2a, M2b) should provide relatively similar
measurements of local relative $A_{\rm J}$ changes. This is because the J-H
and H-K reference values have a much weaker effect on $A_{\rm J}$ differences
than on their absolute values. Figure~\ref{fig:M2ab_Taurus} shows the
correlations between the M2a and M2b extinction maps in a $10\degr \times
10\degr$ region in Taurus. The two maps have an offset of $\Delta A_{\rm
J}=-0.12$\,mag relative to each other but the slope of the linear correlation
is close to one. The slope of an unweighted least squares fit is 0.98 but a
slope 1.0 gives an almost perfect fit up to the highest $A_{\rm J}$ values.
The small non-linearity (the least squares fit underestimating the values at
the highest extinction) is caused by the difference of the M2a and M2b
reference colours that changes as a function of latitude. However, above
$|b|\sim 10\degr$ the effect is already almost unnoticeable, apart from the
zero point differences that makes the appearance of the two all-sky maps
quite different near the plane.

\section{Discussion}  \label{sect:discussion}

We have used the 2MASS survey stars to calculate all-sky NICER and NICEST
maps at 3.0$\arcmin$, 4.5$\arcmin$, and 12.0$\arcmin$ resolution, all sampled
onto Healpix pixels with NSIDE=2048 (pixel size $\sim 1.7\arcmin$). We
discuss below the possible uses of these maps and make comparisons with other
all-sky extinction maps and detailed extinction mappings of selected clouds.

\subsection{All-sky extinction estimates} 

The main motivation of this work was to provide a resource that can be easily
checked for the presence of Galactic extinction. The maps do not and cannot
measure the full extinction along the LOS and are instead weighted
towards nearby structures.

There are previous estimates of extinction over the whole sky. Some of these
``extinction'' maps are, however, in fact derived from dust emission and are
therefore measuring the total amount of dust along the LOS
\citep{Schlegel1998, Planck2014_XI}. First all-sky NIR extinction maps were
presented by \citet{Dobashi2011_2MASS}, together with an extensive catalogue
of dark clouds detected in extinction. Those extinction maps were calculated
based on J-H and H-K colours separately. The main improvement of the new
NICER maps is that it makes an optimised combination of the colour
information contained in the J, H, and K band magnitudes. This should result
in an increase in S/N. Furthermore, the NICEST maps should give a more
unbiased picture of the highest column density structures. By providing
extinction maps in Healpix format, we hope that they will be useful for
studies of extended regions of the sky or in studies involving large number
of targets distributed over the whole sky.

We calculated NICER maps with three alternative definition of the reference
colours that would correspond to zero extinction (see
Sects.~\ref{sect:reference}-\ref{sect:br}). These were M1 (average statistics
at high latitudes, $|b|>60\degr$) and two versions, M2a and M2b, based on
Besan\c{c}con model of the Galactic distribution of stars with different
intrinsic colours (M2a using the intrinsic colours, M2b colours further
reddened by a simple model of diffuse extinction). The resulting extinction
maps showed strong differences at large scales, especially near the Galactic
plane. In spite of this, they give almost identical results for differential
measures of extinction, if these are calculated over fields no larger than a
few degrees. Near the Galactic plane the zero point of the extinction scale
is not well defined because of the unknown mixture of stars and dust along
the LOS. This ambiguity could be resolved only with knowledge of the
3D distribution of the stars and the ISM.

\begin{figure}
\includegraphics[width=8.8cm]{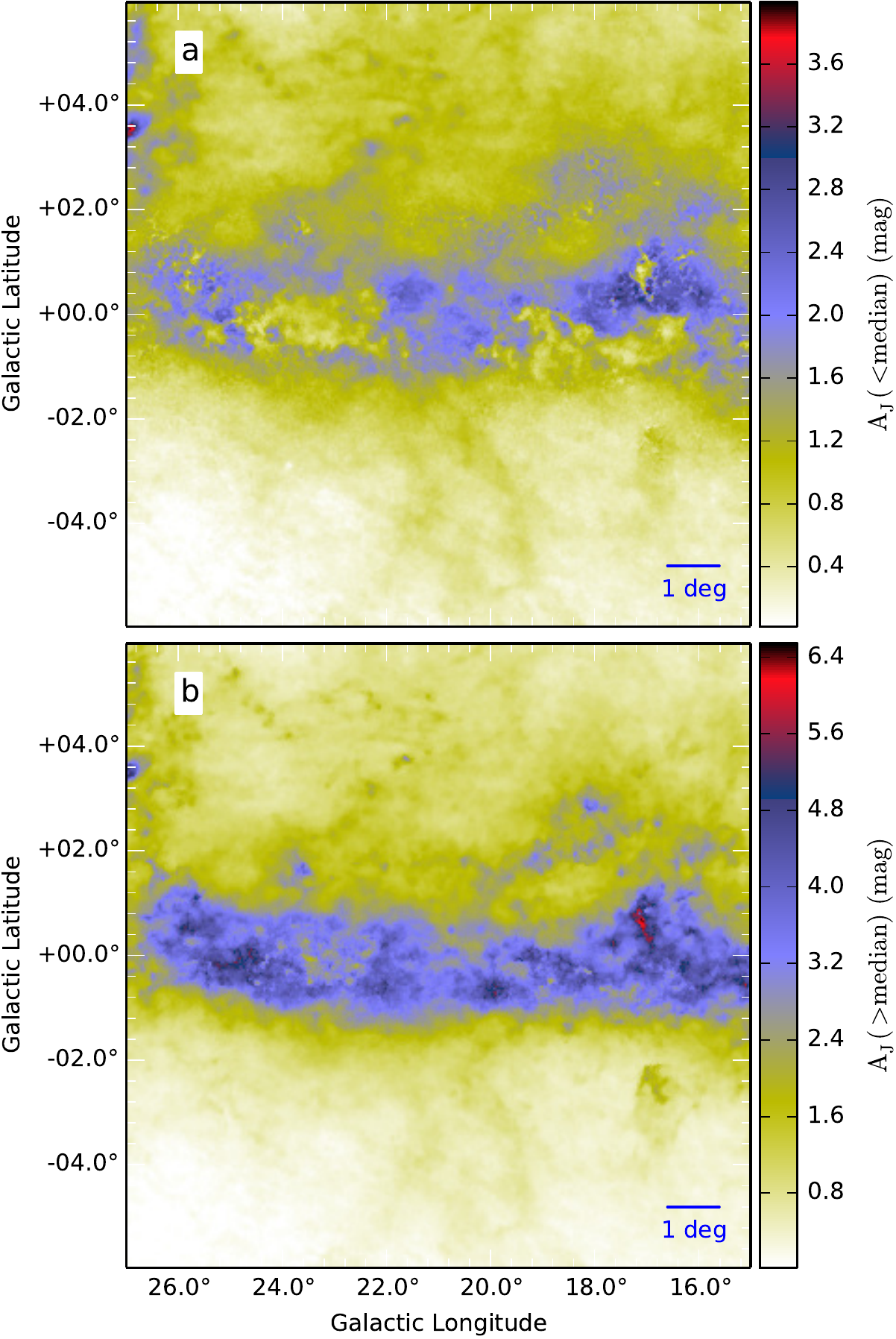} 
\caption{
Comparison of extinction maps using stars below (frame $a$) or above
(frame $b$) the median extinction of all stars within the beam.
Figures show a $6\degr \times 6\degr$ field centred at Galactic
coordinates ($l$, $b$)=(21.0, +0.0) degrees. The spatial resolution of
the maps is 3$\arcmin$ and we have applied offsets to set the minimum
of both maps at $A_{\rm J}\approx 0$\,mag.
}
\label{fig:MOD_plane}
\end{figure}

We believe the all-sky maps to be useful in many studies, at least as the
first step towards more in-depth analysis of the local stellar populations
and the reddening of their colours. The NICER maps (M1, M2a, M2b) and NICEST
map (M1 version) are available on the
web\footnote{http://www.interstellarmedium.org/Extinction}.
The site also includes NICER maps
computed based on the stars below and above the median $A_{\rm J}$ of all
stars in the beam. These may be useful in estimating the robustness of the
extinction estimates. In the Galactic plane, they may also reveal distinct
structures that might be located at different distances along the
LOS (see Fig.~\ref{fig:MOD_plane}). However, to actually prove the
distance differences, more work is needed on the actual 3D distribution of
the dense clouds \citep[cf][]{Marshall2009}.

\subsection{Comparison with other NICER and NICEST maps} \label{sect:other}

We compared our all-sky maps with some previously published extinction maps.
As shown in Sect.~\ref{sect:results}, in a small field the different versions of
extinction maps (M1, M2a, M2b) are very similar, with the exception of the
$A_{\rm J}$ zero point. Therefore, we carry out the comparison using the M2a
map only.

\begin{figure}
\includegraphics[width=8.8cm]{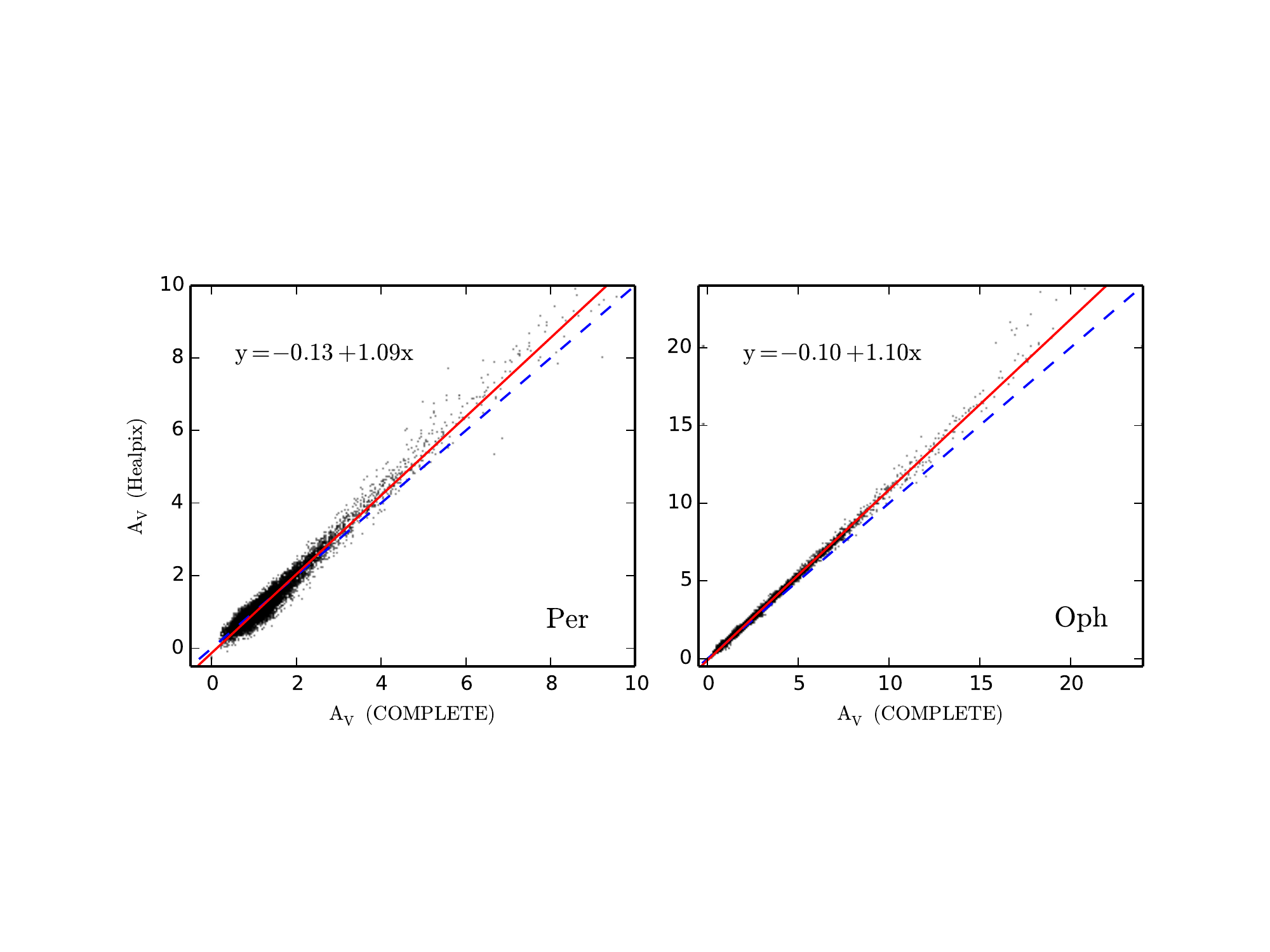} 
\caption{
Comparison at 6$\arcmin$ resolution of the M1 maps and the NICER maps
of Perseus and Ophiuchus regions available at the COMPLETE web site.
In Ophiuchus a half degree radius region centred at ($l$,
$b$)=(351.9, 15.1) is affected by artefacts and is omitted from the
plot. 
}
\label{fig:COMPLETE}
\end{figure}

The COMPLETE data releases include a $9\times 12$ degree NICER extinction map
of the Perseus region, with a pixel size of 2.5$\arcmin$ and a spatial
resolution of 5.0$\arcmin$\footnote{COMPLETE team, ``2MASS Final Perseus
Extinction Map'', http://hdl.handle.net/10904/10080 V2}, and a $9\times 8$
degree NICER map of Ophiuchus, with a spatial resolution of 3$\arcmin$ and a
pixel size of 1.5$\arcmin$\footnote{COMPLETE team, ``2MASS Final Ophiuchus
Extinction Map'', http://hdl.handle.net/10904/10081 V2}.
Figure~\ref{fig:COMPLETE} compares these with our M2a map. We resampled our
data onto the pixels of the COMPLETE maps, and, to reduce the scatter
resulting from the resampling, all data were convolved to a final spatial
resolution of 6.0$\arcmin$.

The overall correspondence is good, the fitted least squares lines having a
slope $\sim$1.1 in both fields. The $\sim$10\% difference in the scaling can
be explained by the adoption of a slightly different extinction curves
(transformation from $A_{\rm J}$ to $A_{\rm V}$). 
Part of the scatter in Fig.~\ref{fig:COMPLETE} can be attributed to the
different resolution of the original data and the uncertainty of the
convolution for maps sampled at steps near the actual resolution. The
relation of M2a vs. COMPLETE is well fitted by a linear relation up to the
highest extinctions. In the correlation with M1 map (not shown) the slopes
are identical but the scatter is somewhat smaller. The residual rms values
were 0.13\,mag and 0.11\,mag instead of 0.18\,mag and 0.13\,mag (Perseus and
Ophiuchus, respectively). This suggests that part of the dispersion
originates because the reference colours in the M2a map change slowly
across the field while they were probably kept constant in the case of the
COMPLETE maps. However, the convolution of our data (originally at
3.0$\arcmin$ resolution but sampled only on $\sim 1.7\arcmin$ pixels) stands
for most of the scatter in Fig.~\ref{fig:COMPLETE}.

\begin{figure}
\includegraphics[width=8.8cm]{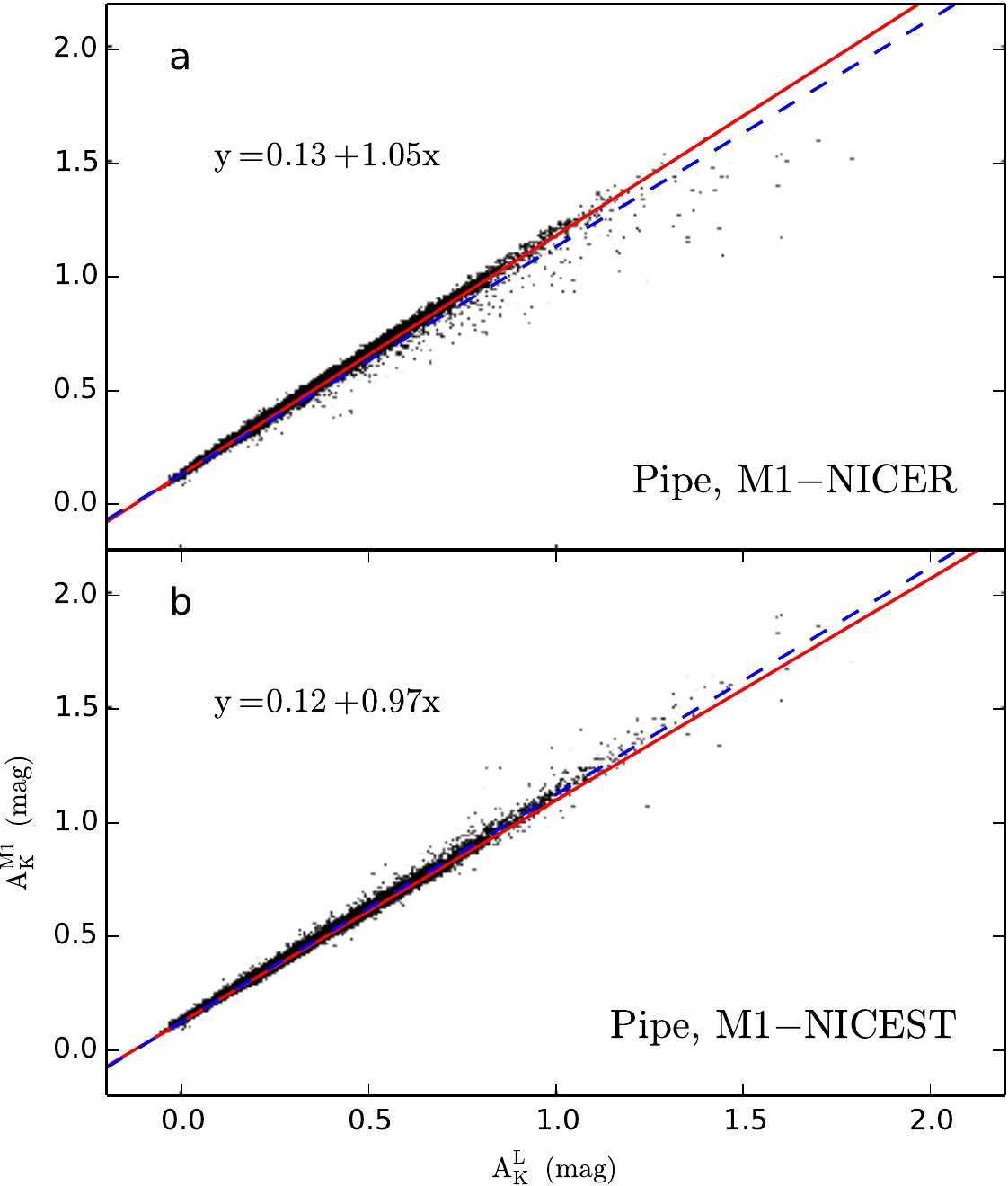} 
\caption{
Comparison at 3$\arcmin$ resolution of our M1 NICER map and the NICER map
($A_{\rm K}^{\rm L}$) of \cite{Lombardi2006_Pipe} convolved from 1$\arcmin$
to the same resolution. Shown are the pixel-by-pixel correlations using
either the M1-NICER (frame $a$) or M1-NICEST map (frame $b$). The pixel size
is 1$\arcmin$.
}
\label{fig:Pipe_new}
\end{figure}

NICER and NICEST extinction maps of the Pipe Nebula are available on the
web\footnote{Lombardi, Marco; Alves, Joao; Lada, Charles J., 2014, ``2MASS
extinction map of the Pipe Nebula'', http://dx.doi.org/10.7910/DVN/25112
Harvard Dataverse Network V3} \citep[see][]{Lombardi2006_Pipe}. These maps
have a resolution of 1.0$\arcmin$ and a pixel size of 30$\arcsec$. 
Figure~\ref{fig:Pipe_new} shows the comparison between this NICER map
(``reference maps'', convolved to 3.0$\arcmin$ resolution) and our M1 maps
calculated with NICER and NICEST methods. The zero point difference of
$\sim$0.12\,mag is again related to the different assumption of the reference
colours. Because the comparison is using NIR extinction $A_{\rm K}$, the
scale should not be sensitive to the adopted extinction law.

The overall match is good but, compared to the reference map, our NICER
estimates are systematically lower at high $A_{\rm K}$. This remains true
even relative to a fit with slope forced to the value of one. Our map was
calculated directly at the 3.0$\arcmin$ resolution and, therefore, suffers
from higher bias because there are more column density variations within the
beam. Using the NICEST method, most of the bias disappears. Thus, these
errors indeed appear to be caused by small scale structures. Compared to the
reference map, the overall scatter is reduced also in regions of lower
extinction. The rms values of the difference are 0.015 and 0.011\,mag, for
Fig.~\ref{fig:Pipe_new} $a$ and $b$, respectively, significantly smaller than
in the case of the Perseus and Ophiuchus fields because
the comparison was done using directly our values on the original Healpix
pixels, carrying out convolution only on the reference data that were well
sampled.

\begin{figure}
\includegraphics[width=8.8cm]{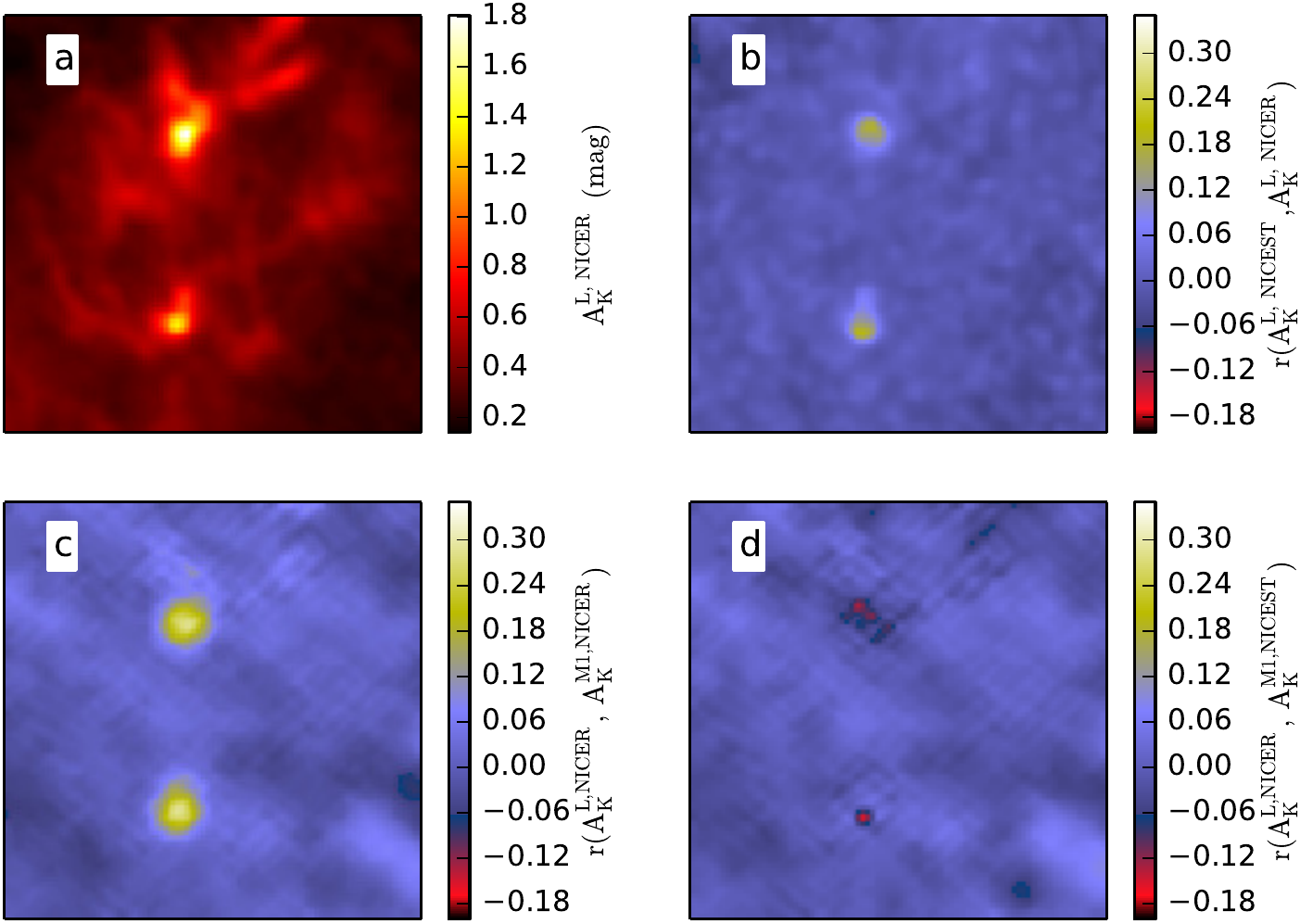} 
\caption{
A detail of the Pipe extinction maps. Frame $a$ shows the 1$\arcmin$
resolution NICER reference map \citep{Lombardi2006_Pipe} convolved to
$3\arcmin$ resolution. The other frames show relative differences between:
Lombardi et al. NICEST and NICER maps (frame $b$; both convolved from
1$\arcmin$ to $3\arcmin$), M1-NICER with the original resolution of
$3\arcmin$ and Lombardi et al. NICER convolved to the same resolution (frame
$c$), and M1-NICEST with the original resolution of $3\arcmin$ and Lombardi
et al. NICER convolved to the same resolution (frame $d$). 
}
\label{fig:Pipe_ratios}
\end{figure}

In the comparison, it is useful to look also at individual structures in the
maps. Figure~\ref{fig:Pipe_ratios} shows a detail of the Pipe extinction map
with two compact clumps. The comparison of the NICER and NICEST reference
maps shows that NICEST method recovers (when examined at 3.0$\arcmin$
resolution, frame $b$) almost 20\% higher peak values of extinction.
Similarly, our M1-NICER map, being calculated at 3$\arcmin$ resolution
instead of the 1$\arcmin$ resolution, shows peak values that are lower by a
similar amount (frame $c$).
M1-NICEST is much closer to the peak values of the reference NICER map. This
shows that NICEST is able to compensate for much of the bias (almost all of
the bias between 1$\arcmin$ and 3$\arcmin$ maps). In fact,
Fig.~\ref{fig:Pipe_ratios}$d$ shows some negative residuals at the location
of the extinction peaks. Thus, our NICEST estimates calculated at 3$\arcmin$
resolution are remarkably between the NICER and NICEST estimates calculated
at 1$\arcmin$ resolution. Of course, even NICEST cannot replace all the
information that is lost in the initial spatial averaging of the extinction
estimates of individual stars. In particular, if some high column density
regions are completely void of background stars, calculated extinction
estimates are still only a lower limit of the true peak extinction.

In Fig.~\ref{fig:Pipe_ratios}c and d some hatch pattern can be seen. This is
because, unlike in Fig.~\ref{fig:Pipe_new}, we have resampled our Healpix
maps onto the pixels of the reference maps. Because the original M1 maps are
defined on Healpix pixels with a size of ~1.7$\arcmin$, after convolution the
residual maps show artificial small scale structure. It would be better to
resample the higher resolution data onto the pixels of the lower resolution
data (as in Fig.~\ref{fig:Pipe_new}) or, preferably, to have better sampling
of all data. However, at the next higher resolution, NSIDE=4096, the files
would four times larger. This would diminish the practical usability of the
maps, with little practical improvement. For the lower resolutions of
4.5$\arcmin$ and 12$\arcmin$, the sampling is more adequate.

\section{Conclusions} \label{sect:conclusions}

We have calculated all-sky extinction maps with 2MASS catalogue stars and
using both NICER and NICEST methods. Calculations have been repeated with
different assumptions of the reference colours corresponding to zero
extinction. The analysis of the extinction estimates and comparison with
extinction maps already available for some regions, we arrive at the
following conclusions.
\begin{itemize}
\item The appearance of the all-sky extinction map depends on the
assumptions of reference colours, which are not well defined when stars and
dust are intermixed along the LOS. However, the relative extinction
estimates always agree very well when calculated for fields no larger than a
few degrees in size.
\item The comparison carried out with extinction maps published for Taurus,
Perseus, Ophiuchus, and Pipe Nebula regions shows good agreement. This
suggests that extinction structures can be mapped relatively precisely
without an in-depth analysis of the stellar populations in the field.
\item Because of small-scale column density variations, NICER maps
underestimate at highest column densities the beam-averaged extinction.
However, comparison with higher resolution maps showed the NICEST method to
be effective in reducing this bias.
\end{itemize}

\begin{acknowledgements}
This publication makes use of data products from the Two Micron All Sky Survey, which is a
joint project of the University of Massachusetts and the Infrared Processing and Analysis
Center/California Institute of Technology, funded by the National Aeronautics and Space
Administration and the National Science Foundation.
MJ acknowledges the support of the Academy of Finland Grant No. 250741.
MJ acknowledges the Observatoire Midi-Pyrenees (OMP) in Toulouse for its
support for a 2 months stay at IRAP in the frame of the `OMP visitor
programme 2014', during which time this work was initiated.
\end{acknowledgements}

\bibliography{biblio_v1.3}

\end{document}